\documentclass[10pt,twocolumn,twoside]{IEEEtran}

\usepackage[utf8]{inputenc} 
\usepackage[T1]{fontenc}
\usepackage{url}
\usepackage{ifthen}
\usepackage{cite}
\usepackage[cmex10]{amsmath} % Use the [cmex10] option to ensure complicance
\usepackage{amssymb} 
\usepackage{amsthm}
\usepackage{tikz}   
\usepackage{mathtools}  
\usepackage{caption}
\usepackage{comment}

\usepackage{graphics,
           psfrag,
           epsfig,
           cite,
           amssymb,
           url,
           dsfont,
           subfigure,
           algorithm,
           algorithmic,
           balance,
           enumerate,
           color,
           setspace
}
\usepackage{amsmath}%[centertags][tbtags][sumlimits][nosumlimits][intlimits][namelimits][nonamelimits]

\usepackage{epstopdf}

\newtheorem{example}{Example}
\newtheorem{definition}{Definition}

\newtheorem{lemma}{Lemma}

\newtheorem{corollary}{Corollary}
\newtheorem{proposition}{Proposition}

\newtheorem{remark}{Remark}
\newtheorem{construction}{Construction}

\DeclarePairedDelimiter\ceil{\lceil}{\rceil}
\DeclarePairedDelimiter\floor{\lfloor}{\rfloor} 
\usepackage{pifont}
 
\newcommand{\xmark}{\text{\ding{55}}}
\usepackage{array}
\newcolumntype{M}[1]{>{\centering\arraybackslash}m{#1}}
\newcolumntype{N}{@{}m{0pt}@{}}

\newcommand{\eref}[1]{(\ref{#1})}

\newcommand{\fref}[1]{Figure~\ref{#1}}

\newcommand{\pref}[1]{Proposition~\ref{#1}}
\newcommand{\cref}[1]{Constraint~\ref{#1}}

\newcommand{\lref}[1]{Lemma~\ref{#1}}

\title{\Large \bf
On the Achievability Region of Regenerating Codes for Multiple Erasures}
 \author{ \IEEEauthorblockN{Marwen Zorgui, Zhiying Wang }\\
 \IEEEauthorblockA{Center for Pervasive Communications and Computing (CPCC)  \\ University of California, Irvine, USA
 \\ \{mzorgui,zhiying\}@uci.edu
 }
  }
\usepackage{stfloats}
\begin{document}
\maketitle
\thispagestyle{empty}
\pagestyle{empty}

\begin{abstract}
We study the problem of centralized exact repair of multiple failures in distributed storage. We describe  constructions that achieve a new set of interior points under exact repair. The constructions build upon the layered code construction by Tian et al in \cite{tian2015layered}, designed for exact repair of single failure. We firstly improve upon the layered construction for general system parameters. Then, we extend the improved construction to support the repair of multiple failures, with varying number of helpers. In particular, we prove the optimality of one point on the functional repair tradeoff of multiple failures for some parameters. Finally, considering minimum bandwidth cooperative repair (MBCR) codes as centralized repair codes, we determine explicitly the best achievable region obtained by space-sharing among all known points, including the MBCR point. 
\end{abstract}
%\vspace{-.4cm}
\section{Introduction}
Driven by the growth of data-centric applications, efficient data storage and retrieval has become of crucial importance for several service providers. Distributed storage systems (DSS) are  currently widely employed for large-scale storage. DSS provide scalable storage and high level of resiliency in the face of server failures. To maintain the desired level of failure tolerance, DSS utilize a replacement mechanism for out-of-access nodes, known also as the repair mechanism, that allows to recover the content of inaccessible/failed nodes. The repair process of a failed node
is performed by downloading data from accessible nodes (or a subset thereof) in
the system and recovering the lost data. Efficiency of a DSS is determined by two parameters, namely, the overhead required for reliability and the amount of data being transferred for a repair process. The seminal work in \cite{dimakis2010network} proposed a new class of erasure codes, called regenerating codes, that optimally solve the repair bandwidth problem. It is shown in \cite{dimakis2010network} that one can significantly reduce the amount of bandwidth required for repair and the bandwidth decreases as each node stores more information. Regenerating codes, as presented in \cite{dimakis2010network}, achieve \textit{functional repair}. In this case, the replacement nodes are not required to be exact copies of the failed nodes, but the repaired code should satisfy reliability constraints. However, in practice, it is often more desirable to recover the exact same information as the failed node, which is called \textit{exact repair}. Exact repair codes are easier to implement and maintain, and thus are of more interest.

There has been a flurry of interest in designing exact repair regenerating codes  \cite{shah2012interference,suh2011exact,rawat2016progress,goparaju2016minimum,cadambe2011optimal,Zigzag_Codes_IT, MSR_optimal_bandwidth,elyasi2016determinant,zorgui_isit_17}. Moreover, there is a growing literature focused on understanding the fundamental limits of exact repair regenerating codes \cite{sasidharan2014improved,duursma2014outer,sasidharan2016outer,duursma2015shortened}, as opposed to the well-understood functional regenerating codes \cite{dimakis2010network}. 
\subsection{Multi-node recovery}
In many practical scenarios, such as in large scale storage systems, multiple failures are more frequent than a single failure. Moreover, many systems apply a lazy repair strategy, which seeks to limit the repair cost of erasure codes. Indeed, it has been demonstrated that jointly repairing multiple failures reduces the overall bandwidth compared to repairing each failure individually \cite{kermarrec2011repairing,closed_form_cooperative_regene,rawat2016centralized,allerton_multi_node}. We distinguish between two ways of repairing multiple failures.

\textit{Cooperative regenerating codes}:  In this framework, each replacement node first downloads information from $d$ nodes (helpers). Then, the replacement nodes exchange information between themselves before regenerating the lost nodes. Of interest to our work, we note that codes corresponding to the extreme points on the cooperative tradeoff have been developed: minimum storage cooperative regenerating (MSCR) codes \cite{li2014cooperative,closed_form_cooperative_regene} and minimum bandwidth cooperative regeneration (MBCR) codes\cite{wang2013exact}. 

\textit{Centralized regenerating codes}: Upon failure of $e$ nodes, the repair is carried out in a centralized way by contacting any $d$ helpers out of the $n-e$ available nodes, $d \le n-e$, and downloading $\beta$ amount of information from each helper. 
The content of any $k$ out of $n$ nodes in the system is sufficient to reconstruct the entire data. Let $\alpha$ be the size of each node, and $F$ be the size of the entire data.
A code satisfying the centralized repair constraints is referred to as an $(
F, n, k, d, e,\alpha, \beta)$ code. We also say it is a code of the $(n,k,d,e)$ system. In our previous work \cite{allerton_multi_node}, we characterized the functional repair tradeoff for multi-node recovery. 
Let $q=\ceil{\frac{k}{e}}-1, t= k-  q e$. 
%It follows that $0<t \le e$, with $t=e \iff e \mid k$ and $t= k \mod e \iff e \nmid k $.
The normalized functional tradeoff can then be written as follows

\vspace{-.4cm}
\begin{small}
\begin{align}
\min(t \bar{\alpha}, d \bar{\beta}) + \sum\limits_{p=0}^{q-1} \min (e \bar{\alpha}, (d-t-p e) \bar{\beta}) \geq 1,
\label{functional_bound}
\end{align}
\end{small}

\vspace{-.4cm}
\noindent where $\bar{\alpha}=\alpha/F, \bar{\beta}=\beta/F$.
%The number of piece-wise linear segments is given by $q=\ceil{\frac{k}{e}}-1$. 
Inequality \eref{functional_bound} gives $q$ linear bounds: 

\vspace{-.5cm}
\begin{small}
\begin{align} 
(t+ p e) \bar{\alpha}+ \sum\limits_{i=p}^{q-1} (d-t-i e) \bar{\beta} \geq 1,\quad p=0,\ldots,q-1.
\label{linear_bounds}
\end{align}
\end{small}

\vspace{-.4cm}
In this work, we are interested in designing \textit{centralized exact repair} regenerating codes for recovering multiple failures. 
When $e \geq k$, the tradeoff reduces to a single point, which is trivially achievable \cite{allerton_multi_node}. We hereafter focus on the case $e<k$. 
%The number of segments is one when $q=1 \iff \ceil{\frac{k}{e}}=2 \iff e < k \le 2e.$

In \cite{rawat2016centralized}, it is argued that cooperative regenerative codes can be used to construct centralized repair codes. The total bandwidth in this case is obtained by taking into account the bandwidth obtained from the helper nodes and disregarding the communication between the replacement nodes. In particular, MSCR codes achieve the same performance as centralized minimum storage multi-node repair (MSMR) codes \cite{rawat2016centralized,zorgui_isit_17}. Additionally, MBCR codes can be used as centralized repair codes, which do not correspond to centralized minimum bandwidth codes on the functional tradeoff \cite{allerton_multi_node}. These points are given by  

\vspace{-.4cm}
\begin{small}
\begin{align}
(\bar{\alpha}_{MSMR},\bar{\beta}_{MSMR})&=(\frac{1}{k}, \frac{e}{k(d-k+e)}) ,\label{MSMR_k_is_d}\\
(\bar{\alpha}_{MBCR},\bar{\beta}_{MBCR})&= (  \frac{2d+e-1}{k(2d-k+e)},\frac{2e}{k(2d-k+e)} ).
\label{MBCR_k_is_d}
\end{align}
\end{small}
 
\vspace{-.4cm}
\noindent\textbf{Contributions of the paper:}
%The main contributions of this paper are summarized as follows.
We  improve upon the layered construction presented in \cite{tian2015layered}, which is concerned with single node repair, to construct a family of regenerating codes that is capable of repairing multiples failures. In particular, for the $(k+e,k,k,e)$ system, we first prove the optimality of a particular constructed point using the functional repair tradeoff;  combining the achievable points via our construction and also the MBCR point, we then characterize the best achievable region obtained by space-sharing between all known points.
 
 The remainder of the paper is organized as follows. A
description of our first code construction is provided in Section II. In Section III, we analyze the achievability region of the $(k+e,k,k,e)$ system. We describe our second code construction in Section IV, before concluding in Section V.

Notation: we denote by $[i]$ the set of integers $\{1,2,\dots,i\}$ for $i \ge 1$.
%%%%%%%%%%%%%%%%%%%%%%%%%%%%%%%%%%%%%%%%%%%%%%%%%%%%%%%%%%%%%%%%%%%%%%%%%%%%%%%%%%%%%%%%%%%%%%%%%%%%%%%%%%%%%%%%%%%%%%%%%%%%%%%%%%%%%%%%%%%%%%%%%%%%%%%%%%%%%%%%%%%%%%%%%%%%%%%%%%%%%%%%%%%%%%%%%%%%%%%%
\section{Code Construction}
Exact repair regenerating codes are characterized
by parameters $(F,n,k,d,e,\alpha,\beta)$. We consider a distributed
storage system with $n$ nodes storing $F$ amount of information. The data elements are distributed across the $n$ storage nodes such that each node can store up to $\alpha$ amount of
information. We use $\bar{\alpha}=\alpha/F, \bar{\beta} = \beta/F$ to denote the normalized storage size and repair bandwidth, respectively. The system should satisfy the following two
properties:

\noindent$\bullet$	\textit{Reconstruction property}: by connecting
to any $k \le n$ nodes it should be sufficient to reconstruct the
entire data.

\noindent$\bullet$	\textit{Repair property}: upon failure of $e$ nodes, a central
node is assumed to contact $d$ helpers, $k \le d \le n-e$, and
download $\beta$ amount of information from each of them.
The exact content
of the failed nodes is determined by the central node. $\beta$ is called the
repair bandwidth. 

We first describe the code construction which is an improvement upon \cite{tian2015layered}. The construction is based on a collection of subsets of $[n]$, called a Steiner system. Information is first encoded within each subset, and then distributed among the $n$ nodes. We recall the definition of Steiner systems. 
\begin{definition}
A  Steiner system $S(t,r,n)$, $t \le r \le n$, is a collection of subsets of size $r$, included in $[n]$, such that any subset of $[n]$ of size $t$ appears exactly once across all the subsets.
\end{definition}
Steiner systems do not exist for all design parameters. When $t=r$, Steiner systems always exist, and the blocks in this case are all $r-$combinations of the set $[n]$.
The family of $(F,n,k,d,e,\alpha,\beta)$ codes we describe below is parameterized by $t,m,r$, for $e \le m < r  \le n, t \le r$, where
%\begin{itemize}
%\item	$n$: the number of storage nodes.
%\item $m$ is related to the number of failures that the system can tolerate,
%\item $r$ represents the number of code symbols per block.
%\end{itemize}
\begin{align}
F=N (r-m),  N= \frac{\binom{n}{t}}{\binom{r}{t}},\alpha= \frac{N r}{n}, k=n-m.
\label{block_design}
\end{align}

\begin{construction} \label{cnstr1}
\textbf{Precoding step:}
We consider a Steiner system $S(t,r,n)$  and generate $N=\frac{\binom{n}{t}}{\binom{r}{t}}$ blocks such that each block is indexed by a set $J \in S(t,r,n)$. Block $J$ corresponds to $r-m$ information symbols over an alphabet of size $q$, which is then encoded using an MSMR code with length $r$ and dimension $r-m$ over an alphabet of size $q$. The codeword symbols, called the repair group $J$, is comprised of $ \{ c_{x,J}: x \in J \}$.  Moreover, we assume that the MSMR code possesses the optimal repair bandwidth \eqref{MSMR_k_is_d} for any number of erasures $l$, $1 \le l \le m$, and any number of helpers $d$,  $r-m \le d \le r-l$. The total number of information symbols is $F= N (r-m)$.

\textbf{The code matrix:}
The code structure can be described by a code matrix $C$, of size $n\times N$. The rows of $C$ are indexed by integers in $[n]$, corresponding to the different storage nodes, and its columns are indexed by sets in $S(t,r,n)$, arranged in some arbitrary chosen order. We formally define $C$ as
\begin{align}
C_{x,{J}}= \begin{cases}
   c_{x,J},& \text{if } x \in {J},\\
    -,              & \text{otherwise},
\end{cases}
\label{code_matrix}
\end{align}
where $"-"$ denotes an empty symbol. Node $i \in [n]$ stores all the non-empty symbols of row $i$ in the code matrix $C$. It can be checked that the storage per node is given by $\alpha= \frac{N r}{n}$.
 \end{construction}
By abuse of notation, the terms block and repair group are used interchangeably. The requirement on the alphabet size $q$ is dictated by the existence of an MSMR code with the required property in \eref{MSMR_k_is_d}. Such MSMR codes are known to exist \cite{MSR_optimal_bandwidth}. 
\begin{example}
\label{code_example}
Consider a Steiner system  $S(t,r,n)=S(3,4,8)$. So the number of blocks is $N=14$ and each node number appears $\alpha=\frac{r N}{n}=7$ times in the blocks. The 14 blocks are given by 

\vspace{-.4cm}
\begin{footnotesize}
\begin{align*}
J_1&=\{1,2,4,8\},
J_2=  \{2,3,5,8\},
J_3=  \{   3,4,6,8\},
J_4=  \{  4,5,7,8\}\\
J_5&=  \{   1,5,6,8\},
J_6=  \{  2,6,7,8\},
J_7=  \{   1,3,7,8\},
J_8=  \{ 3,5,6,7\}\\
J_9&=  \{1,4,6,7  \},
J_{10}=   \{ 1,2,5,7\},
J_{11}=  \{ 1,2,3,6\}\\
J_{12}&=  \{  2,3,4,7\},
J_{13}=  \{1,3,4,5\},
J_{14}=  \{2,4,5,6\}.
\end{align*}
\end{footnotesize}
%\vspace{-.4cm}
\noindent The code matrix is given by (\ref{example_code}) in Figure 1.
%Page \pageref{fig:example_code}.
\begin{figure*}  [b]
\begin{small}
\begin{align}
\label{example_code}
C=\begin{bmatrix}
c_{1,J_1} & -& -&-&c_{1,J_5} &-&c_{1,J_7} &-&    c_{1,J_9}&c_{1,J_{10}}&c_{1,J_{11}}&-&c_{1,J_{13}}&-&\\ 
c_{2,J_1} & c_{2,J_2} &-& -&  - &c_{2,J_6}  &-  & - & - &c_{2,J_{10}}&c_{2,J_{11}}&c_{2,J_{12}}&-&c_{2,J_{14}}\\
-&c_{3,J_2} & c_{3,J_3} &-&-&-& c_{3,J_7} &c_{3,J_8}  &-&-& c_{3,J_{11}}&c_{3,J_{12}}&c_{3,J_{13}}&-&\\
c_{4,J_1} & - &c_{4,J_3}& c_{4,J_4} &-&-&-&-& c_{4,J_9} &-&-& c_{4,J_{12}} & c_{4,J_{13}} & c_{4,J_{14}}\\
-&c_{5,J_2} &-&c_{5,J_4} & c_{5,J_5} & -&-&c_{5,J_8} & -&c_{5,J_{10}} & -&-&c_{5,J_{13}} & c_{5,J_{14}} \\
-&-&c_{6,J_3} & -&c_{6,J_5} & c_{6,J_6} &-& c_{6,J_8} & c_{6,J_9} &-& c_{6,J_{11}} &-&-& c_{6,J_{14}} \\
-&-&-&c_{7,J_4} & -&c_{7,J_6} & c_{7,J_7} & c_{7,J_8} & c_{7,J_9} & c_{7,J_{10}} & -&c_{7,J_{12}} &-&-&\\
c_{8,J_1}&c_{8,J_2} & c_{8,J_3} & c_{8,J_4} & c_{8,J_5} & c_{8,J_6}&c_{8,J_7}&-&-&-&-&-&-&- 
\end{bmatrix}.
\end{align}
\end{small}
\label{fig:example_code}
\caption{Code matrix for the system with parameters $(t,r,n)=(3,4,8)$. }
\end{figure*}
Let $m=2, e=2, d=n-e=6$. Then we can repair nodes 1 and 2 simultaneously, by downloading 

\noindent$\bullet$	symbols $c_{4,J_1}, c_{8,J_1}$ from nodes 4 and 8, respectively. These help repair symbols $c_{1,J_1}$ and $c_{2,J_1}$,

\noindent$\bullet$	symbols $c_{5,J_{10}}, c_{7,J_{10}}$ from nodes 5 and 7, respectively. These help repair symbols $c_{1,J_{10}}$ and $c_{2,J_{10}}$,

\noindent$\bullet$	symbols $c_{3,J_{11}}, c_{6,J_{11}}$ from nodes 3 and 6, respectively. These help repair symbols $c_{1,J_{11}}$ and $c_{2,J_{11}}$,

\noindent$\bullet$	$\frac{1}{2}$ symbol from each of the nodes $5,6$ and $8$, to repair $c_{1,J_5}$,

\noindent$\bullet$	$\frac{1}{2}$ symbol from each of the nodes $3,7$ and $8$, to repair $c_{1,J_7}$,

\noindent$\bullet$	$\frac{1}{2}$ symbol from each of the nodes $4,6$ and $7$, to repair $c_{1,J_9}$,

\noindent$\bullet$	$\frac{1}{2}$ symbol from each of the nodes $3,4$ and $5$, to repair $c_{1,J_{13}}$,

\noindent$\bullet$	and similarly for node 2 to repair $c_{2,J_2},c_{2,J_6},c_{2,J_{12}}$ and $c_{2,J_{14}}$.
 
In total, we download $18$ symbols. Each helper transmits $3$ symbols.
\end{example}
From the example above, we see that each repair group $J$ tolerates the failure of $m$ nodes. Therefore, the code $C$ also tolerates the failure of up to any $m$ nodes. Thus, it can be checked that for Construction \ref{cnstr1} from any $k=n-m$ nodes, we can recover the data, which is the reconstruction parameter. Moreover, the code can recover from any $m$ failures. Therefore, it is possible to repair simultaneously any $1\le e   \le m $ failures. The number of helpers is flexible, and satisfies $ k \le d \le n-e$. The repair bandwidth is given in Propositions \ref{prop:repair_multi} and \ref{prop_steiner} for two different scenarios.
\begin{proposition} \label{prop:repair_multi}
Using Construction \ref{cnstr1} with $t=r$, it is possible to repair simultaneously any set of $1\le e\le m$ nodes, using $n-m \le d \le n-e$ helpers, such that the contribution of each helper, denoted by $\beta_e(d)$, is given by

\vspace{-.4cm}
\begin{footnotesize}
\begin{align}
\label{repair_beta}
&\beta_e(d)=\nonumber\\
& \sum\limits_{s=1}^e \binom{e}{s} \sum\limits_{p=\max(s,r-d)}^{\min(n-d-e+s,r-1)} \binom{d-1 }{r-p-1} \binom{n-d-e }{p-s} \frac{s}{m-p+s}.
 \end{align}
\end{footnotesize}
\end{proposition}
\begin{IEEEproof}
In the repair procedure, any subset of missing symbols belonging to the same repair group is repaired via MSMR repair procedure, using \textit{all} available helpers from the same group among the chosen helper nodes. Fixing the set of helper nodes, we argue that the repair is feasible.
Indeed, let $H$ be the set of $d$ helpers. For each repair group $J$, we denote the set of remaining nodes in $J$ as $J^{'}$. Using $|H \cup J^{'} | \le n-e$ and $d \geq k=n-m$, it follows that
\begin{align}
|J^{'} \cap H | &= |H |+ |J^{'}| -|H \cup J^{'} | \nonumber\\ &\geq d + r-e - (n-e)=r+d-n
\nonumber\\ &
 \geq r-n + n-m=r-m.
\label{intersect}
\end{align}
Thus, for each repair group, we have enough information across the set of helpers to recover the missing components. \\
We now analyze the contribution of a single helper $h$: $h$ helps in the simultaneous repair of $s$ missing symbols of the same repair group, such that $1 \le s \le e$. For each size $s$, we count all possible cases in which the repair can be done through the help of $r-p$ coded symbols among all the $d$ helpers, because the number of available coded symbols determines the contribution of each helper, as dictated by the MSMR repair bandwidth \eqref{MSMR_k_is_d}. It follows that, for the corresponding repair group, $r-p-1$ can be chosen from the set of $d-1$ helpers (helper $h$ already belongs to the repair group by assumption), while the remaining $p-s$ elements of the repair group can be chosen from the remaining $n-e-d$ nodes. \fref{repair_situation} summarizes the repair situation for given parameters $s$ and $p$. Summing over all repair contributions, and analyzing the limit cases of $p$ for a given $s$, \eref{repair_beta} follows.
\end{IEEEproof}
\begin{figure} [H] 
\includegraphics[width=1\linewidth]{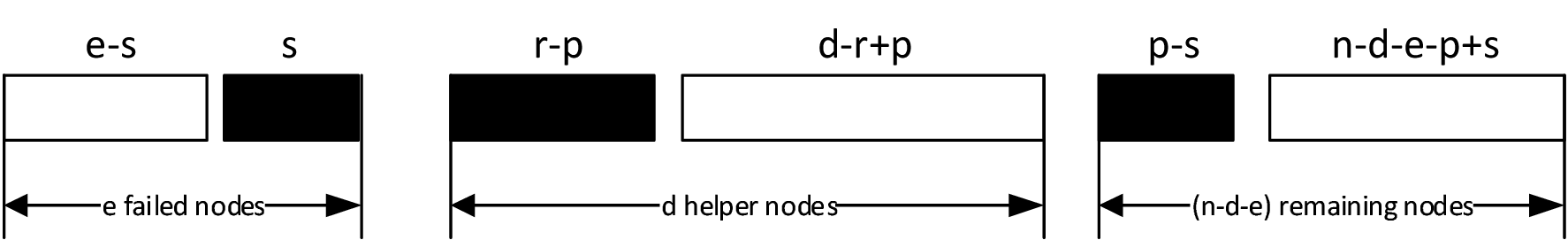}
\caption{{\small A repair situation associated to given parameters $s$ and $p$.} 
%\vspace{-.8cm}
}
\label{repair_situation}
 \end{figure}

\begin{center}
\begin{figure}[H]
\center
\includegraphics[width=1\linewidth]{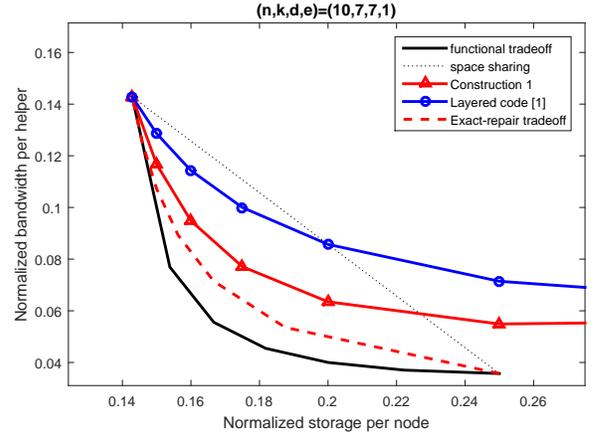}
\caption{Using the MSR repair property improves upon the layered code repair performance.}
\label{comparison_d_equal_k}
\end{figure}
\end{center}

\begin{remark}
\label{benegits_MSMR} It can be seen that the repair procedure can benefit from the MSMR repair property in the case $n>k+1$. In particular, the advantages of using MSMR codes in our construction over maximum distance separable (MDS) codes as in \cite{tian2015layered} are: 1) lower repair bandwidth, 2) symmetric repair among helper nodes, which obviates the need for the expensive procedure of duplicating the block design as in \cite{tian2015layered}, and 3) adaptability, meaning non-trivial repair strategies for multiple erasures, $1\le e \le m$ with the help of varying number of helpers $d$, such that $n-m\le d  \le n-e$.
\fref{comparison_d_equal_k} shows a comparison between the performance of the layered code in \cite{tian2015layered} and Construction \ref{cnstr1}, for an $(n,k,d)=(10,7,7)$ system. The MSR repair property clearly helps reduce the bandwidth.
\end{remark}

The technique of using MSMR codes as building blocks for outer code constructions has been used in the literature, for instance in constructing codes with local regeneration \cite{kamath2014codes,rawat2013optimal}.

\begin{remark}We argue that one can use a regenerating code corresponding to an interior point instead of an MSMR code as the inner code per repair group. Consider the case $(n,k,d,e)=(5,4,4,1)$. Let $r=t=5,m=1$ in \eref{block_design}. The code structure is given by 
\begin{align}
C  =\begin{bmatrix}
%J_1=\{2,3,4,5 \}  & J_2=\{1,3,4,5 \} & J_3=\%{1,2,4,5 \}   & J_4=\{1,2,3,5 \} & J_5=\%{1,2,3,4 \}\\
 - & c_{1,J_2} & c_{1,J_3} & c_{1,J_4} & c_{1,J_5}\\ 
 c_{2,J_1} &  - & c_{2,J_3} & c_{2,J_4} & c_{2,J_5}\\ 
c_{3,J_1} & c_{3,J_2} & -  & c_{3,J_4} & c_{3,J_5}\\ 
c_{4,J_1} & c_{4,J_2} & c_{4,J_3} &  - & c_{4,J_5}\\ 
c_{5,J_1} & c_{5,J_2} & c_{5,J_3} & c_{5,J_4} & -\\ 
 \end{bmatrix}.
\end{align}
Thus, the code per column of $C$ is of length $r=4$ and dimension $ r-m= 3$. We use the interior code: $(\bar{\alpha}_0,\bar{\beta}_0)=(\frac{3}{8},\frac{1}{4})$ per repair group. Let $F_0$ be the information size per column. Thus, $F=5 F_0$ and $\alpha=\frac{3F_0}{2}$. It follows that $ \bar{\alpha}=\frac{3}{10}$. To repair node 1, we download a total bandwidth of $3 F_0$. Thus, $\bar{\beta}=\frac{3}{20}$. We obtain the achievable point $(\bar{\alpha},\bar{\beta})=(\frac{3}{10}, \frac{3}{20})$. The same point is equally achievable using Construction \ref{cnstr1} with $(t,r,n,m,e)=(3,3,5,1,1)$ with an MSMR code as the interior code. This point is optimal on the exact-repair tradeoff of the $(5,4,4,1)$ system \cite{tian2015note,elyasi2016determinant}, and is the optimal point next to the minimum bandwidth regenerating point.
\end{remark}

%\subsection{ General Steiner Systems for an (k+2,k,k,2) system }
In Proposition \ref{prop:repair_multi}, we considered Construction \ref{cnstr1} with Steiner systems such that $t=r$. We study next the use of a general Steiner system for the specific $(k+2,k,k,2$) system.

\begin{proposition} \label{prop_steiner}
Construction 1 generates an $(F,k+2,k,k,2,\alpha,\beta)$ code such that
\begin{align}
\label{Steiner_system_points}
&F=(r-2)\frac{\binom{n }{t }}{\binom{r }{t }} , \alpha=  \frac{\binom{n-1}{t-1}}{\binom{r-1}{t-1}}, \beta=\frac{\binom{n-2}{t-2}}{\binom{r-2}{t-2}}, \\
&  \bar{\alpha}= \frac{r}{n(r-2)}, \bar{\beta}= \frac{r(r-1)}{n(n-1)(r-2)}.
\label{Steiner_system_points_normalized}
\end{align}
\end{proposition}

\begin{IEEEproof}
We consider a Steiner system $S(t,r,n)$  and let $m=2$. From \eref{block_design}, we obtain $F$ and $\alpha$ as in \eref{Steiner_system_points}. To analyze the repair bandwidth per helper, we distinguish two cases:
 
\noindent\textbf{Case $t=2$}: If the helper node $h$ shares a block with both failed nodes, then, by design, $h$ does not share any other block with either of the failed nodes. Thus, $h$ contributes a single symbol ($\log_2 q$ bits) that is useful for the repair of the missing symbols of the shared repair group. Otherwise, $h$ shares exclusively two blocks with each of the failed nodes. In each of the shared repair group, node $h$ contributes $\frac{1}{2}$ symbol ($\frac{1}{2}\log_2 q$ bits) to help repair the corresponding missing symbol, by virtue of the MSMR repair property (i.e., the missing symbol is repaired with $r-1$ helpers). 

\noindent\textbf{Case $t\geq 3$}: For a helper $h$, the number of blocks he shares with both failed nodes is given by 
$
\lambda_3\triangleq \frac{\binom{n-3}{t-3}}{\binom{r-3}{t-3}}.
$
The number of blocks node $h$ shares exclusively with either of the failed nodes is given by 
$
\lambda_2-\lambda_3 \triangleq \frac{\binom{n-2}{t-2}}{\binom{r-2}{t-2}}-\frac{\binom{n-3}{t-3}}{\binom{r-3}{t-3}}.
$
Therefore, the contribution of each helper node is 
\begin{align*}
\beta&=  \frac{2\lambda_3 }{(r-2)-(r-2)+2}+  \frac{2(\lambda_2-\lambda_3)}{(r-1)-(r-2)+1}
=  \lambda_2.
\end{align*}
\end{IEEEproof}

The repair in Example \ref{code_example} is an illustration of Proposition \ref{prop_steiner}. Similar to Proposition \ref{prop:repair_multi}, the repair bandwidth is identical among the  helper nodes, and independent of the choice of the failed nodes and helpers.

\begin{remark}
We note here that $\bar{\alpha}, \bar{\beta}$ do not depend on $t$ by \eref{Steiner_system_points_normalized}. The advantage of using Steiner systems with smaller $t$, whenever they exist, is that they induce smaller $\alpha$ and $\beta$, for the same normalized parameters. Indeed, it can be shown that $\alpha$, as given by \eref{Steiner_system_points}, is strictly increasing in $t$. Therefore, to reduce the storage size per node, and therefore the repair bandwidth, it is advantageous to use a Steiner System with the smallest $t$, $t \le r$. Moreover, when $e=2, t=r$, \pref{prop:repair_multi} and \pref{prop_steiner} give the same $\bar{\alpha}, \bar{\beta}$.
\end{remark}
\section{Analysis of the achievability for an $(n,k,d,e)=(k+e,k,k,e)$ system}
In this section, we analyze the achievable region for an $(n,k,d,e)=(k+e,k,k,e)$ system by means of Construction \ref{cnstr1}, using, for simplicity, a Steiner system with $t=r$.
\begin{proposition}
Construction \ref{cnstr1} with $t=r,m=e$ generates a set of achievable points for an $(F,k+e,k,k,e,\alpha,\beta)$ system, such that

\vspace{-.4cm}
\begin{small}
\begin{align}
\label{achievable_points}
F &= \binom{k+e }{r }(r-e)   , \alpha=  \binom{k+e-1}{r-1} , \beta=\binom{k+e-2 }{r-2} , 
\\  
& \bar{\alpha}= \frac{r}{(k+e)(r-e)}, \bar{\beta}= \frac{r(r-1)}{(k+e)(k+e-1)(r-e)} , \nonumber\\ & \quad e+1 \le r \le k+e.
\label{achievable_points_bar}
\end{align}
\end{small}
\end{proposition}
\begin{IEEEproof}
When $d=k, n=k+e, m=e$, $r$ is chosen such that $e+1\le r \le n$, the general expression in \eref{repair_beta} is given by

\vspace{-.4cm}
\begin{small}
\begin{align}
\beta_e(k)
%&= \sum\limits_{t=\max(1,r-k)}^{\min(e,r-1)} \binom{e}{t}   \binom{k-1 }{r-t-1}  \frac{t}{e } \nonumber\\
&= \sum\limits_{t=\max(1,r-k)}^e \binom{e-1 }{t-1}  \binom{k-1 }{r-t-1} =\binom{k+e-2 }{r-2},
\label{pre_Vandermonde}
\end{align}
\end{small}

\vspace{-.4cm}
\noindent where the last equality follows from Vandermonde's identity.
\end{IEEEproof} 
\subsection{Optimality of one achievable point}
\begin{proposition}
\label{optimal_point}
For the $(k+e,k,k,e)$ system, the point achieved in \eref{achievable_points} for $r=k+e-1$ is an optimal interior point.
\end{proposition}
\begin{IEEEproof}
From \eref{achievable_points} when $r=k+e-1$, we achieve $F=(k+e)(k-1), \alpha=k+e-1, \beta=k+e-2$. Thus, 
\begin{align}
\label{achievable_point}
(\bar{\alpha},\bar{\beta})=(\frac{k+e-1}{(k+e)(k-1)}, \frac{k+e-2}{(k+e)(k-1)}).
\end{align}
Substituting \eref{achievable_point} in \eref{linear_bounds} and setting $p=q-1$, we obtain  
\begin{align*}
(t+q e -e) \bar{\alpha} + (d- t - q e +e)\bar{\beta}= (k-e)\bar{\alpha}+ e \bar{\beta}= 1.
\end{align*}
Therefore, the above point lies on the functional repair lower bound and hence is optimal. It lies on the first segment of the bound near the MSMR point, and it is not the MSMR point nor the MBCR point, as indicated by \eref{MSMR_k_is_d} and \eref{MBCR_k_is_d}.
\end{IEEEproof}

\subsection{Optimal extension property}
From \pref{optimal_point}, Construction \ref{cnstr1} gives us an optimal point for any $(k+e,k,k,e)$ system. Construction \ref{cnstr1} also offers the following property.
\begin{proposition}
\label{optimal_extension}
Consider a $(k+e,k,k,e)$ system and consider the optimal point achieved by Construction 1 in \pref{optimal_point}, one can \textit{extend} the system to a $(k+e+1,k,k,e+1)$ system, operating at the optimal point of \pref{optimal_point}, by adding another node to the system and increasing the storage per node, while keeping the initial storage content.
\end{proposition}
\begin{IEEEproof}
Let $\alpha_i,\beta_i,F_i$, for $i=1,2$, refer to the parameters of the old and the new systems, respectively. Then, $\alpha_2-\alpha_1=1, \beta_2-\beta_1=1, F_2-F_1=k-1$. Moreover, the number of blocks $N$ is increased by 1. Let $k+e+1$ be the index of the new node to be added. The new code is obtained by simply adding another block, whose set is $\{1,\ldots,k+e \}$, and adding to the old sets the element $(k+e+1)$ to each of them, and thus generating another coded symbol for the corresponding repair group. A key requirement is to assume the use of an MSMR code that can accommodate the addition of extra coded symbols, when needed. This can be done by choosing the number of nodes of the MSMR code to be as large as needed (this may result in an increase in the underlying field size). Each old node will store an extra symbol coming from the new repair group, while the new node stores the newly generated coded symbols from the old repair groups. 
\end{IEEEproof}
\begin{example}
We illustrate the process of extending a $(4,3,3,1)$ system to a $(5,3,3,2)$ system. Initially, each repair group is of size 3. The code blocks are given by 
\begin{align*}
J_1&=\{2,3,4  \}  , J_2=\{1,3,4  \} , J_3=\{1,2,4  \}   , J_4=\{1,2,3 \}.
\end{align*}
The code matrix is given by

\vspace{-.35cm}
\begin{small}
\begin{align*}
C_1=\begin{bmatrix}
 - & c_{1,J_2} & c_{1,J_3} & c_{1,J_4}  \\ 
 c_{2,J_1} &  - & c_{2,J_3} & c_{2,J_4}  \\ 
c_{3,J_1} & c_{3,J_2} & -  & c_{3,J_4}  \\ 
c_{4,J_1} & c_{4,J_2} & c_{4,J_3} &  -  
 \end{bmatrix}.
\end{align*}
\end{small}

\vspace{-.35cm}
\noindent Adding node 5 to the system, we add another block $J_5=\{1,2,3,4 \}$, whose symbols will be distributed across the old nodes $\{1,2,3,4 \}$. The old blocks become

\vspace{-.35cm}
\begin{small}
\begin{align*}
J_1=\{2,3,4,5 \}  , J_2=\{1,3,4,5 \} , J_3=\{1,2,4,5 \}   ,J_4=\{1,2,3,5 \}.
\end{align*}
\end{small}

\vspace{-.35cm}
\noindent The new node  $5$ stores newly generated coded symbols of each of the old repair groups $\{J_1,\ldots,J_4 \}$. The new code matrix is given by
\begin{align*}
C_2  =\begin{bmatrix}
%J_1=\{2,3,4,5 \}  & J_2=\{1,3,4,5 \} & J_3=\%{1,2,4,5 \}   & J_4=\{1,2,3,5 \} & J_5=\%{1,2,3,4 \}\\
 - & c_{1,J_2} & c_{1,J_3} & c_{1,J_4} & c_{1,J_5}\\ 
 c_{2,J_1} &  - & c_{2,J_3} & c_{2,J_4} & c_{2,J_5}\\ 
c_{3,J_1} & c_{3,J_2} & -  & c_{3,J_4} & c_{3,J_5}\\ 
c_{4,J_1} & c_{4,J_2} & c_{4,J_3} &  - & c_{4,J_5}\\ 
c_{5,J_1} & c_{5,J_2} & c_{5,J_3} & c_{5,J_4} & -\\ 
 \end{bmatrix}.
\end{align*}
\end{example}
The above property is useful for systems for which the fault tolerance may be deemed insufficient. Therefore, one can increase the fault tolerance of the system without sacrificing the optimality on the exact repair tradeoff, or changing the existing data. We note also that by a successive application of \pref{optimal_extension}, we can increase the fault tolerance of the system by any desirable factor.

\subsection{Acheivability region for the $(k+e,k,k,e)$ system}
In this subsection, we seek to determine the convex hull of the known achievable points for the $(k+e,k,k,e)$ system. The convex hull, denoted by $\mathcal{R}$, is the smallest convex set containing all known achievable points, obtained by all convex combinations (i.e., space-sharing) among the points achieved by Construction \ref{cnstr1}, described in \eref{achievable_points_bar}, and also the MBCR point given by \eref{MBCR_k_is_d}. The objective is therefore to determine which points are sufficient to describe $\mathcal{R}$. We refer to these points as \emph{corner points} of $\mathcal{R}$. 

\begin{figure} 
\begin{center}
\includegraphics[width=1\linewidth]{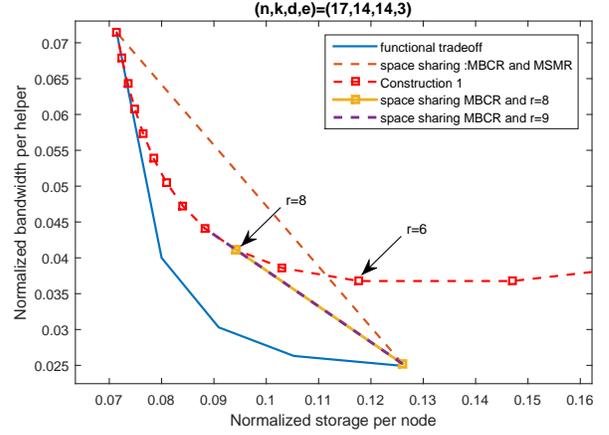}
\caption{Achievable points for an $(n,k,d,e)=(17,14,14,3)$ system. The x-axis is the normalized storage per node  $\bar{\alpha}$ and the y-axis is the normalized bandwidth $\bar{\beta}$.}
\label{convex_hull}
\end{center}
 \end{figure}

\fref{convex_hull} presents the achievable points for an $(17,14,14,3)$ system. 
The achievable points of \eref{achievable_points} are parameterized by $r$, such that $e+1 \le r \le e+k$. For each $r$, we denote the corresponding point as $(\bar{\alpha}_{r },\bar{\beta}_{r })$. As $r$ decreases, the storage $\alpha_r$ increases. By abuse of notation, we refer to the point $(\bar{\alpha}_{r },\bar{\beta}_{r })$ as point $r$. 
%Note that the MBCR point does not lie on the functional tradeoff unless XXXXXXXXXXX (in \fref{convex_hull} the gap is small).
We state some guiding observations for our subsequent analysis. First, one can eliminate some of the achievable points obtained by Construction \ref{cnstr1}. For instance, point $r=5$, with $\bar{\alpha}=0.1471$, achieves a similar bandwidth as the neighbor point $r=6$, but at a larger storage size. Points to the right of $\bar{\alpha}=0.1471$, such that $r < 5$, can be also immediately eliminated, because they can be outperformed by space-sharing between the MBCR point and some interior point. Interestingly, we observe that point $r=8$ lies exactly on the segment joining point $r=9$ and the MBCR point. This means that, while point $r=8$ is not outperformed by space-sharing, it is nonetheless not necessary for the description of $\mathcal{R}$, and thus it is not considered as a corner point. In the following, we show that the observations from \fref{convex_hull} can be generalized and we explicitly determine the corner points of $\mathcal{R}$, depending on the system's parameters $e$ and $k$.
\begin{lemma}
\label{limiting_r_range}
The achievable points in \eref{achievable_points}, with $r< 2 e$, are not corner points in $\mathcal{R}$.
\end{lemma}
 \begin{IEEEproof}
From \eref{achievable_points}, it can be seen that $\bar{\alpha}(r)$, seen as a function of $r$, is decreasing. $\bar{\beta}(r)$ is a fractional function in $r$, with a pole at $r=e$. For $r>e$, $\bar{\beta}(r)$ is convex in $r$. It can be shown that it decreases and then increases monotonically. Therefore, as $\bar{\alpha}(r)$ is decreasing, the points of interest are those for which $\bar{\beta}(r)$ increases. Moreover, by noticing that $\bar{\beta}(2e)=\bar{\beta}(2e-1)$, it follows that points with $r \le 2e-1$ do not contribute to the acheivability region $(\bar{\alpha},\bar{\beta})$, as these points are outperformed by the point $r=2e$ in terms of both, storage and bandwidth.
\end{IEEEproof}
\lref{limiting_r_range} implies that it is sufficient to consider the range $ 2 e \le r \le k+e $. We define  the non-negative integer $p$ such that $r=2e+p$. We now show  that the achievable points $r=2e,\ldots,k+e$ can not be eliminated by space-sharing between themselves, when not considering the MBCR point.
\begin{lemma}
\label{decreasing_slope}
The achievability region of the points $(\bar{\alpha}_r,\bar{\beta}_r), r=e+1,\ldots,k+e$ has points with $r \in \{ 2e,\ldots,k+e\}$ as corner points, when not considering the MBCR point.
\end{lemma}
\begin{IEEEproof}
By virtue of \lref{limiting_r_range}, points with $e+1 \le r <  2e $ can be eliminated. We consider the segment joining the points $(\bar{\alpha}_r,\bar{\beta}_r)$ and $ (\bar{\alpha}_{r+1},\bar{\beta}_{r+1})$. The slope of the segment, denoted $sl(r)$, is given by
\begin{align*}
sl(r)=\frac{\bar{\beta}_{r+1}-\bar{\beta}_r} {\bar{\alpha}_{r+1}-\bar{\alpha}_r}=\frac{-r(r-2e+1)}{e(k+e-1)}.
\end{align*}
The slope $sl(r)$ is strictly decreasing in $r$ for $r \ge 2 e$. This means, for any three consecutive points $ (\bar{\alpha}_{r+2},\bar{\beta}_{r+2})$, $(\bar{\alpha}_{r+1},\bar{\beta}_{r+1})$ and $(\bar{\alpha}_{r },\bar{\beta}_{r })$, the point $ (\bar{\alpha}_{r+1},\bar{\beta}_{r+1})$ lies below the segment joining the other two extreme points. Therefore, space-sharing between $(\bar{\alpha}_{r+2},\bar{\beta}_{r+2})$ and $ (\bar{\alpha}_{r },\bar{\beta}_{r })$ is suboptimal.
\end{IEEEproof}
Now, we analyze the achievability region when adjoining the MBCR point to the points in \eref{achievable_points} with $2e \le r \le k+e$. 
\begin{lemma}
\label{MBCR_corner_point}
The MBCR point is a corner point for $\mathcal{R}$ .
\end{lemma}
\begin{IEEEproof}
Noting that $\bar{\alpha}_{\text{MBCR}}= \frac{2k+e-1}{(k+e)k} > \bar{\alpha}_{2e}=\frac{2}{ (k+e)}$ and $\bar{\beta}_{\text{MBCR}}= \frac{2e}{(k+e)k} < \bar{\beta}_{2e}=\frac{2(2e-1)}{ (k+e) (k+e-1)}$, along with \lref{decreasing_slope} concludes the result.
 \end{IEEEproof}

By \lref{MBCR_corner_point}, we only need to analyze whether space-sharing between the MBCR point and any other point $r$ may outperform some of the other achievable points $r^{'}$.
\begin{lemma}
\label{cross_over}
If a point $r$, $r \ge 2e$, is not outperformed by space-sharing between the point $r+1$ and the MBCR point, then, all points $r'$ such that $r' \geq r$, are corner points of the achievability region. 
\end{lemma}
\begin{IEEEproof}
The assumption of the lemma implies that the slope of the segment joining the points $ (\bar{\alpha}_r, \bar{\beta}_r)$ and $(\bar{\alpha}_{\text{MBCR}}, \bar{\beta}_{\text{MBCR}})$ is smaller than the slope of the segment between  $ (\bar{\alpha}_{r+1}, \bar{\beta}_{r+1})$ and $(\bar{\alpha}_{\text{MBCR}}, \bar{\beta}_{\text{MBCR}})$. As from \lref{decreasing_slope}, the slope of the segment between $(\bar{\alpha}_r,\bar{\beta}_r)$ and $ (\bar{\alpha}_{r+1},\bar{\beta}_{r+1})$ is decreasing in $r$, it follows that no point $r ' \geq r$ can be outperformed by space-sharing across any two other achievable points, including the MBCR point.
\end{IEEEproof}
Therefore, to determine the corner points of $\mathcal{R}$, we need to successively test for increasing values of $p$, such that $0 \le p \le k-e$, whether the point $r=2e+p$ is outperformed by space-sharing of MBCR and point $r+1$. Let $p^{*}$ denote the smallest $p$ such that $r=2e+p$ is not outperformed by space-sharing, it follows by \lref{cross_over} the following  achievability region.

\begin{proposition}
\label{exact_region}
The achievability region $\mathcal{R}$ is given by the corner points

\begin{footnotesize}
\begin{align}
\mathcal{R}= \{ (\bar{\alpha}_r, \bar{\beta}_r) : r \in  \{ r: r=2e+p \text{ and } p^{*} \le p \le k-e\} \cup \{ \text{MBCR} \} \},
\end{align}
\end{footnotesize}

\vspace{-.3cm}
\noindent where $1 \le p^{*} \le k-e$, and $ p^{*}$ is given by
\begin{small}
\begin{align}
 p^*&= \floor*{ \frac{e - k - 2 e^2 + 1+ \sqrt{\Delta} }{2 (e + k - 1)}}+1,\nonumber\\
  \Delta&=(2 e^2 - e + k - 1)^2+8 (  k +e- 1)  e (e - 1) (k-e-1).
 \label{p_star_expanded}
 \end{align}
\end{small}
\end{proposition}
\begin{IEEEproof}
Consider $r=2e+p, 0 \le p \le k-e-1$. We consider space-sharing between the MBCR point and the point $r+1$. We compute the normalized bandwidth, denoted by $\bar{\beta^{'}}_r$, achieved by the considered space-sharing, at the intermediate point $\alpha=\alpha_r$, and then determine whether $\bar{\beta^{'}}_r > \bar{\beta}_r$. Using \eref{achievable_points_bar} and \eref{MBCR_k_is_d}, we obtain after simplification
 
\begin{footnotesize}
\begin{align}
\label{delta_expression}
&\bar{\beta^{'}}_r- \bar{\beta }_r   \nonumber\\
&=\frac{k (- 2 e^2 + 2 e + p^2 + p) - p (- 2 e^2 + e + 1) + 2 e (e^2 - 1) + p^2 (e - 1)}{(e+k)(e+p)(e + k - 1) (e^2 + p e + k - p + k p - 1)}  \nonumber\\
&\triangleq \frac{N_1(k)}{D}\\
&= \frac{(k+e-1)p^2 + p (2 e^2+k-e-1)+2 e (e-1) (e+1-k)}{(e+k)(e+p)(e + k - 1) (e^2 + p e + k - p + k p - 1)} \nonumber\\
&\triangleq \frac{N_2(p)}{D}.
\label{N_2}
\end{align}
\end{footnotesize}

\vspace{-.35cm}
We regard $N_1$ as a function of $k$, for fixed $e$ and $p$, and $N_2$ as a function of $p$, for fixed $e$ and $k$. In this proof, we are interested in analyzing $N_2$. We analyze $N_1$ in a later proof. 

Clearly $D > 0$. Thus, $\text{sign} (\bar{\beta^{'}}_r- \bar{\beta }_r)= \text{sign} (N_2(p))$. Therefore, it suffices to study the sign of $N_2(p)$. We note that $\bar{\beta^{'}}_r- \bar{\beta }_r \le 0$ implies that point $r=2e+p$ can be eliminated by space-sharing and thus it is a not a corner point.  $N_2(p)$ is a quadratic function in $p$. 
%As its leading coefficient is $k+e-1 \geq 0$, then,  $\bar{\beta^{'}}_r- \bar{\beta }_r$ satisfies  $\bar{\beta^{'}}_r- \bar{\beta }_r \le 0$ only if $\Delta \geq 0$ , where 
Let $\Delta$ denote the discriminant of $N_2(p)$. It can be checked that 

\vspace{-.35cm}
\begin{small}
\begin{align*}
\Delta=(2 e^2 - e + k - 1)^2+8 (  k +e- 1)  e (e - 1) (k-e-1) >0.
\end{align*}
\end{small}

\vspace{-.35cm}
\noindent Thus, there exists $p_{0,1},p_{0,2}$ such that $N_2(p_{0,1})=N_2(p_{0,2})=0$. As the leading coefficient of $N_2(p)$ is positive, and $N_2(0)=- 2 e (e-1) (k-e-1)\le 0$, it follows that one solution, say $p_{0,1}$, is negative and the other solution $p_{0,2}$ is non-negative. 
%Both solutions have opposite signs. 
That is, $p_{0,1} <0$ and $p_{0,2} \geq 0$. 
Then, it follows that $ \forall 0 \le p \le p_{0,2}, N_2(p) \le 0$, which implies that the set $\{p: p \le p_{0,2} \}$ can be eliminated. In particular, $p=0$ is always eliminated. Let $p^*= \floor*{p_{0,2}}+1$, as in \eref{p_star_expanded}. Thus, $p^{*}$ outperforms space-sharing and so do all $p \geq p^{*}$. As $N_2(k-e-1)=(k-e)(k+e-1)(k-e-1) \geq 0$, it follows that $p_{0,2}\le k-e-1 $, and thus $ p^* \le k-e$.
%Finally, we note that $N_2(p=k-e-1)=(k-e)(k+e-1)(k-e-1) \geq$. Thus, $p_{0,2}\le k-e-1 $, which implies that $ p^* \le k+e$. The number of points is $k-e+2-p^*$.  
%
%\begin{enumerate}
%\item If $k=e+1$: Then, it can be checked that $N_2(p)=2 e p (e+p)$, which implies that:
%\begin{itemize}
%\item	If $p=0$: then   $N_2(p)=0$. Thus, the point $r=2e$ can be overlooked. Moreover, the point $r=2e$ lies exactly on the line joining MBCR and $r=k+e=n$. $r=2e$ corresponds in this case to the point \eref{optimal_point}. Moreover, in this case, we already know that MBCR point is optimal.
%
%\item	If $p=1$: the point $r=2e+1=k+e$ is the MSMR point and it is a corner point.
%\end{itemize}
%
%\item	If $k>e+1$: then $N_2(p) $ is a quadratic function. 
%It can be checked that $\Delta=(2 e^2 - e + k - 1)^2+8 (  k +e- 1)  e (e - 1) (k-e-1) >0$. 
%
% \end{enumerate}
\end{IEEEproof}
\pref{exact_region} agrees with known particular cases. 1) When $e=1$, we have $p^*=1$ and the only eliminated point $(p=0)$ coincides with the MBCR point, in agreement with \cite{tian2015layered}.
2) The optimal point in \pref{optimal_point} ($p=k-e-1$) is not a corner point for $k=e+1$, because of $p^{*}=k-e>p$ and \pref{exact_region}. Indeed, the point with $p=k-e-1$ lies exactly on the segment joining the MBCR and the MSMR point. 3) When $k>e+1$, the optimal point in \pref{optimal_point} is a corner point, as
$\bar{\beta^{'}}_{k+e-1}- \bar{\beta }_{k+e-1}= \frac{(k-e)(k-e-1)}{k (k+e) (k-1)^2}> 0
$.

While \pref{exact_region} characterizes exactly $\mathcal{R}$, it does not give insight into when a particular point $r=2e+p$ is a corner point or not. We focus on the analysis of the sign of $N_1(k)$ in \eref{delta_expression}. $N_1(k)$ is linear in $k$. Depending on the sign of its the leading coefficient $- 2 e^2 + 2 e + p^2 + p$, there may exist an integer $k_{\text{th}}$ such that when $k \geq k_{\text{th}}$ space sharing enhances the achievability region (i.e., $N_1(k)\le 0$ ) and does not enhance it when $k < k_{\text{th}}$. That is, a point with the same $r$ may be a corner point for some $(k+e,k,k,e)$ systems and may be not a corner point for other systems, with higher reconstruction parameter $k$.

For example, when $p=e-1$, we have $N_1(k)=  e (1-e) (k-5e+1)$. It follows that, for systems with $k \geq 5 e- 1$, the point $r=2e+(e-1)=3e-1$ is outperformed by space-sharing. For systems with $3e-1 \le k <5 e- 1$, the point $r$ is a corner point.

The next proposition addresses the cases in which a particular point $r=2e+p$ is a corner point, using a similar argument as the above example.
\begin{proposition}
\label{r_behavior}
Consider the achievable point $r=2e + p$, for fixed $(e,k), e>1$. Let $ p_{\text{max}}=  \floor*{ \frac{1}{2} (\sqrt{8 e(e-1)-1 }-1)}$ and $k_{\text{th}}=
\ceil*{
 (1-e) \frac{ \binom{p+1}{2}  + 2\binom{e+1}{2}+  e p }{ \binom{p+1}{2} - 2  \binom{e}{2}}}$. Then, Table \ref{tab:contribution} specifies the scenarios in which $(\bar{\alpha}_r,\bar{\beta}_r)$ is a corner point in $\mathcal{R}$. 
%\begin{center}
%\begin{tabular}{|M{2cm}|M{2cm}|M{2cm}|N}
%\hline
%$(\bar{\alpha}_r, \bar{\beta}_r)$ & $k < k_{\text{th}}$ & $k \geq k_{\text{th}}$ &\\[10pt]
%\hline
%$ p \le  p_{\text{max}}$ &  $\checkmark$ &  $\xmark$  &\\[10pt]
%\hline
%$ p >  p_{\text{max}}$ &  $\checkmark$ &  $\checkmark$  &\\[10pt]
%\hline
%\end{tabular}
%\captionof{table}{Summary of cases for which $(\bar{\alpha}_r,\bar{\beta}_r)$ is a corner point in $\mathcal{R}$. The symbol $\checkmark$ means $(\bar{\alpha}_r,\bar{\beta}_r)$ is a corner point while the symbol $\xmark$ denotes the other case.\label{tab:contribution} }
%\end{center}
\begin{center}
\begin{tabular}{|M{2cm}|M{2cm}|M{2cm}|N}
\hline
$(\bar{\alpha}_r, \bar{\beta}_r)$ & $k < k_{\text{th}}(p)$ & $k \geq k_{\text{th}}(p)$ & \\[10pt]
\hline
$ p \le  p_{\text{max}}$ &  $\checkmark$ &  $\xmark$    &\\[10pt]
\hline
$p >  p_{\text{max}}$& \multicolumn{2}{|c|}{$\checkmark$ } &\\[10pt]
\hline
\end{tabular}
\captionof{table}{Summary of cases for which $(\bar{\alpha}_r,\bar{\beta}_r)$ is a corner point in $\mathcal{R}$. The symbol $\checkmark$ means $(\bar{\alpha}_r,\bar{\beta}_r)$ is a corner point while the symbol $\xmark$ denotes the other case.
\label{tab:contribution} }
\end{center}
\end{proposition}

\begin{IEEEproof}
%\begin{remark}
%When $e=1$, we have 
%\begin{align}
%N_1(k)=N_2(p)= k p (p+1)
%\end{align}
%Therefore, by analyzing $N_2(p)$, we immediately see that when $p=0$, we can eliminate the point $p=0$: this exactly the MBCR point. All $p>0$, $N_2(p)>0$, and thus should be retained. This coincides with the known achievability region for an $(k+1,k,k,1)$ system.
%\end{remark}
We examine $N_1(k)$.
First, we note that when $-2e^2+2e + p^2+p >0$, the point $r=2e +p$ is a corner point for all systems. Indeed, as $N_1(e+1)=2e p (e+p) >0, p>0$, we have $N_1(k)> 0, \forall k \geq e+1, p>0$.
It follows that, for a fixed $(e,p)$, we need to determine the sign of $-2 e^2+2e+ p^2+p$. We have
\begin{align}
&-2 e^2+2e+ p^2+p <0 \iff p(p+1) <  2 e(e+1),\\ 
\label{inequality_used_k_0_incresing}
&\iff \binom{p+1}{2} <  2\binom{e}{2},\\
& \iff p < \sqrt{2 e^2-2 e -\frac{1}{4}}-\frac{1}{2} = \frac{1}{2} (\sqrt{8 e(e-1)-1 }-1).
\label{p_UB}
\end{align}
We note that RHS of \eref{p_UB} can not be an integer, as otherwise $\sqrt{8 e(e-1)-1 }$ should be an odd integer, implying $8 e(e-1)-1 \equiv 1 \mod 4$, which leads to a contradiction as $8 e(e-1)-1 \equiv 3 \mod 4$. This also implies that the slope of $N_1(k)$ cannot be 0, for $e>0, \forall p \geq 0$. The maximum value of $p$ satisfying \eref{p_UB} is given by
\begin{align}
p_{\text{max}}=  \floor*{ \frac{1}{2} (\sqrt{8 e(e-1)-1 }-1)}.
\label{p_max} 
\end{align}
Thus, a point $r=2e+p, p > p_{\text{max}}$ is a corner point for any $(k+e,k,k,e)$ system such that $r \le k+e $. For each $0 \le p \le  p_{\text{max}} $, the point $r=2e +p$ is a corner point if and only if $\text{sign} (\bar{\beta^{'}}_r- \bar{\beta }_r )=\text{sign} (N_2(k))> 0$. From \eref{delta_expression}, Let $k_{0}$ be the solution to the linear equation $N_1(k)=0$. Then, after simplification, we have

\vspace{-.35cm}
%\begin{footnotesize}
\begin{align}
k_{0}&= 
%\frac{- 2 e^3 - 2 e^2 p - e p^2 + e p + 2 e + p^2 + p}{- 2 e^2 + 2 e + p^2 + p}
  \frac{(1-e) (2 e^2 + 2 e p + 2 e + p^2 + p)}{- 2 e^2 + 2 e + p^2 + p} \\
& = (1-e) \frac{\binom{p+1}{2} + 2\binom{e+1}{2}+  e p }{ \binom{p+1}{2} - 2  \binom{e}{2}}.
\label{k_0_formula}
\end{align}
%\end{footnotesize}
As $p \le p_{\text{max}}$, we have $- 2 e^2 + 2 e + p^2 + p <0$, which also implies that $k_{0} >0$. As $N_1 (e+1)= 2 e p (e+p)$, we have $k_{\text{th}} \geq e+1$, with equality iff $p=0$. It can be checked from \eref{k_0_formula} that when $p=e-1$, $k_{0}= 5 e-1$. For $k \geq k_0$, point $r$ is not a corner point. As $k$ is an integer and $k_0$ is not necessarily an integer, it follows that $k \geq k_0 \iff k \geq \ceil*{k_0} \triangleq k_{\text{th}}$.

%\noindent\textbf{p=0:} Then, $N_1(k)=2 e (1-e) (k-e-1)$. Thus, for $k=e+1$: the point $(\bar{\alpha}_{2e}, \bar{\beta}_{2e})$ lies exactly on the line joining the MBCR point and $(\bar{\alpha}_{2e+1}, \bar{\beta}_{2e+1})= (\bar{\alpha}_{k+e}, \bar{\beta}_{k+e})$. In this case, $r=k+e-1$ and we already showed in \pref{optimal_point} the optimality of this point. Moreover, we know that the MBCR is optimal in this scenario as $k \equiv 1 (\mod e)$. Therefore, the point $r=2 e$ can be achieved by space-sharing when $k=e+1$. 

%Moreover, for $k> e+1$, $N_1(k)<0$: therefore, space-sharing strictly outperforms the point $r=2e$.

 %\item	Fixing $e$ and $p$: the set  $\{k: k \geq r=2e+p \text{ and } N_1(k) \le 0 \}$ determines the systems for which including the point $(\bar{\alpha}_r, \bar{\beta}_r)$ is not helpful for the achievability region. As $N_1(k)$ is linear in $k$:  (or vice versa, depending on .
 \end{IEEEproof}
Using \pref{r_behavior}, Corollary \ref{cor:second_expression} follows.
\begin{corollary}
\label{cor:second_expression}
For a $(k+e,k,k,e)$ system with $e \geq 2$, we have
\begin{itemize}
\item  $p^{*}$ in \eref{p_star_expanded} can also be expressed as 
\begin{align}
p^{*}&=1 + \max \{ p: p \le p_{max} \text{ and }  k \geq k_{\text{th}}(p) \}\\
&=1 + \max \big\{    p: p \le \floor*{ \frac{1}{2} (\sqrt{8 e(e-1)-1 }-1)} \nonumber\\
& \quad \text{ and }  k \geq \ceil*{
 (1-e) \frac{\binom{p+1}{2} + 2\binom{e+1}{2}+  e p }{ \binom{p+1}{2} - 2  \binom{e}{2}}} \big\}.
\end{align}

\item	The number of corner points in $\mathcal{R}$ is given by $n_c \triangleq |\{ r: 2e+ p^{*} \le r \le k+e\}|+1= k-e+2-p^*$.

\item	As a function of $k$, $p^{*}$ levels out at $k= k_{\text{th}}(p_{max})$ and its final value is given by $ 1+p_{max}$.
\end{itemize}
\end{corollary}

\begin{example}
We consider the setting of \fref{convex_hull}: $e=3,k=14$. We obtain $p_{max}=2, p^{*}=3$. This means the points $r$, for $6 \le r \le 2e+p^{*}-1=8$ are not corner points in $\mathcal{R}$ and the number of corner points is $n_c=10$. This clearly matches the observations made in \fref{convex_hull}.
\end{example}

%%%%%%%%%%%%%%%%%%%%%%%%%%%%%%%%%%%%%%%%%%%%%%%%%%%%%%%%%%%%%%%%%%%%%%%%%%%%%%%%%%%%%%%%%%%%%%%%%%%%%%%%%%%%%%%%%%%%%%%%%%%%%%%%%%%%%%%%%%%%%%%%%%%%%%%%%%%%%%%%%%%%%%%%%%%%%%%%%%%%%%%%%%%%%%%%%%%%%%%%%%%%%%%%%%%%%%%%%%%%%%%%%%%%%%%%%%%%%%%%%%%%%%%%%%%%%%%%%%%%%%%%%%%%%%%%%%%%%%%%%%%%%%%%%%%%%%%%%%%%%%%%%%%%%%%%%%%%%%%%%%%%%%%%%%%%%%%%%%%%%%%%%%%%%%%%%%%%%%%%%%%%%%%%%%%%%%%%%%%%%%%%%%%%%%%%%%%%%%%%%%%%%%%%%%%%%%%%%%%%%%%%
\section{Code construction 2}
In this section, we present another family of codes improved upon \cite{tian2015layered} that encapsulates Construction \ref{cnstr1} as a special case. 

Let $G$ denote the $N(r-m) \times n\alpha$ generator matrix after vectorization of the code in \eref{code_matrix}, with $t=r$. 
Every node corresponds to a set of $\alpha$ columns of $G$. Different from Construction \ref{cnstr1}, we allow $k \le n-m$, hence we may feed $F_c\triangleq (r-m)N$ dependent symbols to the generator matrix.
Let $T$ be $k \alpha$ columns of $G$ corresponding to $k$ out of the $n$ nodes. Let $G|_{T}$ be the submatrix of $G$ consisting of the columns of $T$.
Then the rank of  $G|_{T}$, denoted by $\rho_{k,m,r}$, is
independent of the choice of the $k$ nodes, and is given by \cite{tian2015layered}
\begin{align}
\label{information_size}
\rho_{k,m,r}= \sum\limits_{p=\max(1,r-(n-k))}^{\min(k,r)} \binom{k}{p}\binom{n-k}{r-p} \min(p,r-m).
\end{align}
The maximum amount of information that can be stored in the system, $F$, is upper bounded by $\rho_{k,m,r}$, i.e., $F \le \rho_{k,m,r}$. For instance, when $m=n-k$, it can be checked that $\rho_{k,m,r}= (r-m) N = F_c$. 

To generate the $F_c$ dependent symbols, we add another layer of inner code to Construction \ref{cnstr1}. Moreover, the information symbols are assumed to be over $\mathbb{F}_q^\kappa$, for the finite field $\mathbb{F}_q$ and an appropriately chosen positive integer $\kappa$. 

\begin{construction}\label{cnstr3}
For an $(n,k,d,e)$ system, similarly to Construction \ref{cnstr1}, the code construction is parameterized by $m,r$, such that $e \le m \le n-k$ and $m+1 \le r \le n$ (we assume $t=r$).
For each pair $(r,m)$, let $F$ be given by \eref{information_size}, $\alpha=\binom{n-r}{r-1}$. 
First, the $F$ information symbols $ \{ v_i \}_{i=1}^{F}$, $v_i \in \mathbb{F}_q^\kappa$, are used to construct a linearized polynomial
\begin{align}
f(x)= \sum\limits_{i=1}^{F} v_i x^{q^{i-1}}.
\end{align}
The linearized polynomial
is then evaluated at $F_c $ elements of $\mathbb{F}_q^\kappa$ to obtain $\{f(\theta_i) , 1 \le i \le F_c\}$, which when viewed as vectors over $\mathbb{F}_q$, are linearly independent. 
Finally, the evaluation points $\{f(\theta_i) , 1 \le i \le F_c\}$ are fed to the encoder in Construction \ref{cnstr1}. 

\textbf{Repair:} The repair of $e$ nodes is similar to Construction \ref{cnstr1}, and the contribution of each helper is given by \eref{repair_beta}. 
\end{construction}
We note that the elements in Construction \ref{cnstr1} are defined over an alphabet of size $q$, while the evaluation points are defined over $\mathbb{F}_q^\kappa$. This difference can be resolved by viewing $\{f(\theta_i) , 1 \le i \le F_c \}$ as vectors over $\mathbb{F}_q$ and applying Construction \ref{cnstr1} to each of their components. Similarly, the repair is  carried out component-wise. The linearized polynomial evaluations are an instance of rank-metric codes. In \cite[Proposition 5]{tian2015layered}, it is shown that the use of rank-metric codes guarantees the reconstruction property of the regenerating code. Moreover, \cite{tian2015layered} shows that when $\kappa \ge F_c$, the symbols $\{f(\theta_i) , 1 \le i \le F_c\}$ can be made independent over $\mathbb{F}_q$. In fact, rank metric codes may be replaced by other linear codes, as long as the reconstruction property is satisfied, so as to reduce the field size \cite{tian2015layered}. Furthermore, we note that when $m=n-k$, the use of rank-metric codes is not needed, and the code obtained is simply the code in Construction \ref{cnstr1}.

\begin{remark}
Construction \ref{cnstr3} generalizes the non-canonical construction in \cite{tian2015layered}, which is designed for repairing single erasures. Moreover, the non-canonical construction in \cite{tian2015layered} is based on MDS codes, rather than MSMR codes, and its repair scheme is based on the naive repair scheme of MDS codes. Finally, non-canonical codes in \cite{tian2015layered} set $m=n-d$, while in Construction \ref{cnstr3}, $m$ takes arbitrary values, such that $e \le m \le n-k$.
\end{remark}
 
\begin{center}
\begin{figure} 
\center
\includegraphics[width=1\linewidth]{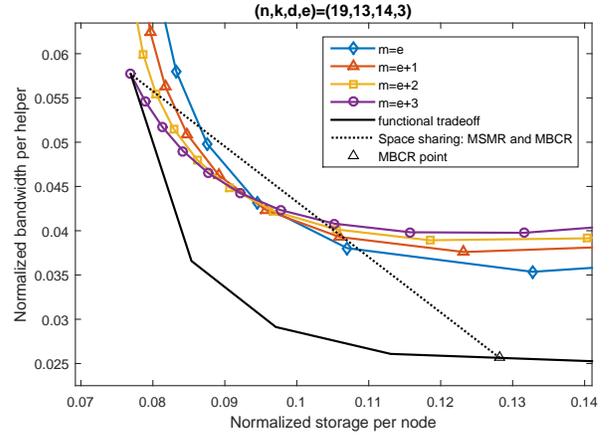}
\caption{Achievable points using Construction \ref{cnstr3} for an $(n,k,d,e)=(19,13,14,3)$ system. The x-axis is the normalized storage per node  $\bar{\alpha}$ and the y-axis is the normalized bandwidth $\bar{\beta}$. When $m=n-k=e+3$, the blue curve coincides with Construction 1.}
\label{multiple_erasures_compare_schemes}
 \end{figure}
 \end{center}

\begin{remark}
The repair process in Construction \ref{cnstr3} does not take into account the dependency introduced by rank-metric codes among the $F_c=(r-m)N$ intermediate symbols. It may be possible to reduce further the repair bandwidth by leveraging such dependency. 
\end{remark}
By varying $m$ and $r$ in Construction \ref{cnstr3}, we obtain various achievability points. Construction \ref{cnstr1} is a special case of Construction \ref{cnstr3}, corresponding to $m=n-k$. In particular, when $k=d, n=k+e$, Constructions \ref{cnstr1} and \ref{cnstr3} coincide as $m=n-k=e$. For other parameters, simulation shows that Construction \ref{cnstr1} performs better closer to the MSMR point while Construction \ref{cnstr3} with $m=e$ performs better closer to the MBCR point. \fref{multiple_erasures_compare_schemes} plots the achievable points by Construction \ref{cnstr3} for an $(n,k,d,e)=(19,13,14,3)$ system, for various values of $m, e \le m \le n-k$. 
%\bl{ Question: when $m=n-k$, if a linearized code of rate 1 is used, do we still need $\kappa \ge F_c$? In this case, it means that we require a large yet unnecessary alphabet.   }
%%%%%%%%%%%%%%%%%%%%%%%%%%%%%%%%%%%%%%%%%%%%%%%%%%%%%%%%%%%%%%%%%%%%%%%%%%%%%%%%%%%%%%%%%%%%
%%%%%%%%%%%%%%%%%%%%%%%%%%%%%%%%%%%%%%%%%%%%%%%%%%%%%%%%%%%%%%%%%%%%%%%%%%%%%%%%%%%%%%%%%%%% 
\section{Conclusion}
We studied the problem of centralized exact repair of multiple failures in distributed storage. 
We first described a construction that achieves a new set of interior points. In particular, we proved the optimality of one point on the functional centralized repair tadeoff. Moreover, considering minimum bandwidth cooperative repair codes as centralized repair codes, we determined explicitly the best achievable region obtained by space-sharing among all known points, for the $(k+e,k,k,e)$ system. Finally, we described another construction, that includes the first construction as a special case, and that generates various achievable points for a general $(n,k,d,e)$ system. Future work includes investigating outer bounds for the centralized exact repair problem. 
%%%%%%%%%%%%%%%%%%%%%%%%%%%%%%%%%%%%%%%%%%%%%%%%%%%%%%%%%%%%%%%%%%%%%%%%%%%%%%%%%%%%%%%%%%%%
%%%%%%%%%%%%%%%%%%%%%%%%%%%%%%%%%%%%%%%%%%%%%%%%%%%%%%%%%%%%%%%%%%%%%%%%%%%%%%%%%%%%%%%%%%%% 
\bibliographystyle{IEEEtran}
\bibliography{biblio2}
\end{document}